\documentclass[sigchi]{acmart}

\usepackage{balance} 

\usepackage{booktabs} 

\copyrightyear{2019}
\acmYear{2019}
\setcopyright{acmlicensed}
\acmConference[CHI 2019]{CHI Conference on Human Factors in Computing Systems Proceedings}{May 4--9, 2019}{Glasgow, Scotland UK}
\acmBooktitle{CHI Conference on Human Factors in Computing Systems Proceedings (CHI 2019), May 4--9, 2019, Glasgow, Scotland UK}
\acmPrice{15.00}
\acmDOI{10.1145/3290605.3300361}
\acmISBN{978-1-4503-5970-2/19/05}

\begin{document}

\title[Applying Dual Systems Theory to Digital Self-Control Tools]{Self-Control in Cyberspace: Applying Dual Systems Theory to a Review of Digital Self-Control Tools}

\author{Ulrik Lyngs}
\email{ulrik.lyngs@cs.ox.ac.uk}
\affiliation{
  \institution{Department of Computer Science, University of Oxford}
  \city{Oxford}
  \state{}
  \country{United Kingdom}
  \postcode{OX1 3QD}
}
\author{Kai Lukoff}
\email{kai1@uw.edu}
\affiliation{
  \institution{Human Centered Design \& Engineering,
University of Washington}
  \city{Seattle}
  \state{}
  \country{USA}
  \postcode{WA 98195}
}
\author{Petr Slovak}
\email{p.slovak@ucl.ac.uk}
\affiliation{
  \institution{Department of Informatics,
King's College London}
  \city{London}
  \state{}
  \country{United Kingdom}
  \postcode{WC1E 6EA}
}
\additionalaffiliation{
  \institution{UCL Interaction Centre, University College London}
  \city{London}
  \state{}
  \country{United Kingdom}
  \postcode{}
}
\author{Reuben Binns}
\email{reuben.binns@cs.ox.ac.uk}
\author{Adam Slack}
\email{AdamSlack@outlook.com}
\affiliation{
  \institution{Department of Computer Science, University of Oxford}
  \city{Oxford}
  \state{}
  \country{United Kingdom}
  \postcode{OX1 3QD}
}
\author{Michael Inzlicht}
\email{michael.inzlicht@utoronto.ca}
\affiliation{
  \institution{Department of Psychology,
University of Toronto}
  \city{Toronto}
  \state{}
  \country{Canada}
  \postcode{M5S 3G3}
}
\author{Max Van Kleek}
\email{max.van.kleek@cs.ox.ac.uk}
\author{Nigel Shadbolt}
\email{nigel.shadbolt@cs.ox.ac.uk}
\affiliation{
  \institution{Department of Computer Science, University of Oxford}
  \city{Oxford}
  \state{}
  \country{United Kingdom}
  \postcode{OX1 3QD}
}

\renewcommand{\shortauthors}{Lyngs et al.}

\begin{abstract}
Many people struggle to control their use of digital devices. However, our understanding of the design mechanisms that support user self-control remains limited. In this paper, we make two contributions to HCI research in this space: first, we analyse 367 apps and browser extensions from the Google Play, Chrome Web, and Apple App stores to identify common core design features and intervention strategies afforded by current tools for digital self-control. Second, we adapt and apply an integrative dual systems model of self-regulation as a framework for organising and evaluating the design features found. Our analysis aims to help the design of better tools in two ways: (i) by identifying how, through a well-established model of self-regulation, current tools overlap and differ in how they support self-control; and (ii) by using the model to reveal underexplored cognitive mechanisms that could aid the design of new tools.
\end{abstract}



\maketitle

\hypertarget{introduction}{%
\section*{Introduction}\label{introduction}}
\addcontentsline{toc}{section}{Introduction}

Smartphones and laptops give their users access to an astonishing range
of tasks anywhere, anytime. While this provides innumerable benefits, a
growing amount of public discussion and research attention focuses on a
perhaps unexpected downside
\citep{Baumer2013, McDaniel2014, Lundquist2014, Wu2016, Alter2017, Leslie2016, Foer2016, Tiku2018}:
Having immense amounts of functionality available instantly and
permanently often makes it difficult for users to focus on their current
task and avoid being overly distracted by notifications or habitual
check-ins \citep{Marotta2017, Thomas2016a, Stothart2015, Dabbish2011}.
This challenge is compounded by the business models of many tech
companies, which incentivise design that nudges people into using
services frequently and extensively in order to optimise advertising
revenue (cf.~the `attention economy'
\citep{Davenport2001, Wu2016, Harris2016, Einstein2016}).

In response, a recent movement in HCI has called for more research into
intentional `non-use' of information and communication technologies
(ICTs) \citep{Baumer2013, Satchell2009, Hiniker2016}. Initial work looked
into why some users quit, or take breaks from Facebook \citep{Baumer2013},
Twitter \citep{Schoenebeck2014}, or other social networking sites, and a
substantial body of related research has now established that a majority
of users feel conflicted about the time they spend with
internet-connected digital technologies and struggle with effective
self-control
\citep{Ames2013, Foot2014, Portwood-Stacer2012, Lee2014c, Sleeper2015, Ko2015, Ko2016, Lundquist2014}.

Researchers within this space have started to design, implement and test
novel tools for supporting self-control over device use, using
interventions such as gamification and social sharing of total time
spent on one's smartphone (with rewards for reducing use) \citep{Ko2015} or
visualisation of laptop use \citep{Whittaker2016}. Meanwhile, an
entire market niche has appeared on the Android and Apple app stores, as
well as on browser extension `web stores', wherein hundreds upon
hundreds of apps and extensions cater to people struggling with
self-control over device use, and provide interventions claimed to help
users. Some of these `digital self-control tools' --- such as \emph{Forest}
\citep{Forestapp.cc2018}, which gamifies self-control through growing of
virtual trees -- have gathered millions of users \citep{GooglePlay2017}.

Yet, while the challenge of supporting self-control over use of
always-connected digital devices has become widely discussed, our
understanding of how best to approach it remains limited \citep[cf.][]{Whittaker2016, Mark2017, Mark2018, Cox2016}. While a growing number
of studies have developed and evaluated novel design interventions, no
systematic reviews have mapped design features in the hundreds of digital
self-control tools that currently exist on the app and web stores \citep[cf.][]{VanVelthoven2018}.
Moreover, new design interventions developed by HCI researchers have
mostly been informed by user interviews and intuitions of interface
designers \citep{Lochtefeld2013, Hiniker2016, Whittaker2016}, or by
theories including cognitive load theory \citep{Collins2014}, Social
Cognitive Theory \citep{Ko2015} and nudge theory \citep{Okeke2018, Kim2016}.
Meanwhile, the dual systems and value-based models prominent in current
cognitive neuroscience research on self-regulation\footnote{In cognitive psychology, \emph{Self-regulation} is commonly used as an umbrella term for all regulatory processes in the service of goal-directed behaviour, including automatic habits, and \emph{self-control} more restrictively about conscious, deliberate attempts at overriding immediate impulses that conflict with one's goals \citep{Hagger2010, Baumeister2007, Milyavskaya2018, Duckworth2014}.}
\citep{Botvinick2015, Vohs2016a, Shenhav2017, Berkman2017, Shea2014, Wood2016}
have yet to be applied \citep[cf.][]{Cox2016, Pinder2018}.

In this paper, we review 367 apps and browser extensions for digital
self-control from the Google Play, Chrome Web, and Apple App stores, and
identify common design features and strategies. As a theoretical lens to
organise and evaluate these tools and provide a deeper understanding of
the self-control challenges they seek to address, we adapt and apply an
integrative dual systems model drawn from established work within the
cognitive neuroscience of self-regulation. Extending recent attempts at
applying dual systems theory to digital behaviour change interventions
\citep{Pinder2018}, our formulation of the model incorporates recent research
on the `expected value of control' \citep{Shenhav2013} as mediator of the
strength of conscious self-control. This, we argue, demystifies the
concept of self-control and helps clarify how specific design features
may work to scaffold successful self-control.

\hypertarget{related-work}{%
\section{Related Work}\label{related-work}}

\hypertarget{evidence-on-self-control-challenges}{%
\subsection{Evidence on Self-Control Challenges}\label{evidence-on-self-control-challenges}}

It has long been known that digital device use for some subset of users
can become associated with severe breakdowns of self-regulation, causing
distress or impaired functioning in important life domains
\citep{Chakraborty2010a}. Thus, for more than two decades, the concept of
`addiction' has been applied by some researchers to such instances,
originally in terms of `internet addiction' \citep{Young1998}, and more
recently `social media' and `Facebook addiction'
\citep{Sleeper2015, Marino2018, Marino2018a, Ryan2014, Andreassen2016}, as well as
`cell phone' and `smartphone addiction'
\citep{Chakraborty2010a, Sapacz2016}.

The current surge of public discussion around self-control struggles and
unwanted distraction - as well as related initiatives by some of the tech giants
\citep{Google2018, Gonzalez2018, Apple2018} - has a broader focus, namely
daily self-control struggles experienced by most users
\citep[cf.][]{Wolwerton2018, Centers2018}. Here, an accumulating body of evidence
suggests that a majority of people experience difficulties with
self-control over device use
\citep{Lundquist2014, Ko2015, Lee2014c, Ko2016, Hiniker2016, Whittaker2016}.
For example, in a survey by Ko et al. \citep{Ko2015}, a majority of
smartphone users felt they were overusing their devices (64\%) and wanted
to change their usage habits (60\%). The patterns respondents wished to
change clustered around two themes: too frequent short usage, where
incoming notifications or urges to e.g.~check the news derailed focus
from tasks they wished to complete; and excessive long usage, where
e.g.~habitually checking devices before bedtime `sucked them in'. Most
users also reported that their strategies for changing this behaviour most
often failed, especially when relying on `willpower', because good
intentions to limit use tended to be overridden by momentary impulses \citetext{\citealp{Elhai2016, Ko2015}; \citealp[cf.][]{Hofmann2012b,Lim2017, Przybylski2013}}.
In addition to a growing amount of evidence from such surveys
\citep{Ko2016, Hiniker2016}, the wealth of articles and opinion pieces that
in recent years have appeared on the topic in major news outlets, viral
blog posts, and popular science books \citep[e.g.][]{Alter2017, Wu2016, Harris2016, Knapp2013, Foer2016, Popescu2018},
as well as the fact that Apple, Google, and Facebook perceive a
sufficiently large demand for features supporting user self-control to
respond \citep{Apple2018, Google2018, Ranadive2018}, suggest that
day-to-day struggles with self-regulation over device use are very
common.

\hypertarget{theory-applied-in-related-research}{%
\subsection{Theory Applied in Related Research}\label{theory-applied-in-related-research}}

\hypertarget{digital-self-control-tools}{%
\subsubsection{Digital Self-Control Tools}\label{digital-self-control-tools}}

To understand the theory use in prior relevant HCI work, we identified 17 HCI papers which have either built novel design intervention or evaluated existing interventions to support self-control over digital device use -- see Table \ref{tab:table-theory-applied-in-digi-self-control}.
For each of these, we examined which self-regulation theories were applied to guide tool development (if designing new interventions) and/or to evaluate effects on user behaviour and perceptions.
These papers were a subset of a large body of literature on self-control in relation to digital device use, spanning psychology, neuroscience, behavioural economics, philosophy, and HCI, continuously collected over a period of two years (500+ papers altogether).
While the literature included here aimed at comprehensively summarising current work in this area of HCI, we note that they were not collected through a formal systematic review process.

In 7 out of 17 papers, no underlying self-regulation theory was specified, with tool development and/or assessment informed only by user-centered design methods such as surveys and interviews with target users and/or design experts.
The 10 remaining papers all referred to distinct theoretical models from psychology (Social Cognitive Theory, classical conditioning, strength model of self-control), cognitive neuroscience (attentional resource theory, cognitive load theory, multitasking and inhibitory brain function), (behavioural) economics (nudge theory, framing effects, rational choice), behaviour change, and addiction research.
For the purposes of the present paper, we note that none of the papers reviewed relied on dual systems models of self-regulation \citep[cf.][]{Pinder2018, Cox2016}.

\begin{table*}
\small

\caption{\label{tab:table-theory-applied-in-digi-self-control}Theories applied in existing work on digital self-control tools.}
\centering
\begin{tabular}{p{1.7cm}p{10cm}p{5.2cm}}
\toprule
Paper & Summary & Self-regulation theory\\
\midrule
Lottridge et al. 2012 \cite{Lottridge2012} & Firefox extension which classifies URLs as work or non-work, then makes non-work tabs less prominent and displays time spent & Multitasking, inhibitory brain function \cite{Gazzaley2008}\\
Löchtefeld et al. 2013 \cite{Lochtefeld2013} & \textit{AppDetox}, an Android app which let users voluntarily create rules intended to keep them from certain apps & None\\
Collins et al. 2014 \cite{Collins2014} & \textit{RescueTime}, a commercial Windows/Mac application which provides visualisations of how much time is spent in different applications & Cognitive Load theory \cite{Block2010}\\
Lee et al. 2014 \cite{Lee2014} & \textit{SAMS}, an Android app for tracking smartphone usage and setting time limits for app use & Relapse prevention model \cite{Witkiewitz2004, Gustafson2011}, clinical guidelines for internet addiction \cite{Young1999}\\
Ko et al. 2015 \cite{Ko2015} & \textit{NUGU}, a smartphone app which let users set goals for limiting usage, then share performance with friends and receive encouragement & Social Cognitive Theory \cite{Bandura1991}\\
Andone et al. 2016 \cite{Andone2016} & \textit{Menthal}, a smartphone app displaying the 'MScore', a single number summarising overall phone usage, as well as a series of main usage measures & None\\
Hiniker et al. 2016 \cite{Hiniker2016} & \textit{MyTime}, an Android app showing time spent in apps (and whether a daily limit was hit) plus a daily prompt asking what the user wished to achieve & None\\
Kim et al. 2016 \cite{Kim2016} & \textit{TimeAware}, an ambient Windows and Mac widget which presents time spent in 'distracting' or 'productive' applications & Framing effects \cite{Marteau1989}\\
Ko et al. 2016 \cite{Ko2016} & \textit{Lock n' LoL}, a smartphone app which lets users as a group set their phones in a lock mode in which notifications are muted and usage restricted & None\\
Ruan et al. 2016 \cite{Ruan2016} & \textit{PreventDark}, an Android app which detects phone use in the dark and notifies the user that they should put it away & None\\
Whittaker et al. 2016 \cite{Whittaker2016} & \textit{MeTime}, a computer application providing a floating visualisation of time spent in different applications within the last 30 mins & None\\
Kim et al. 2017 \cite{Kim2017} & \textit{Let's FOCUS}, an Android and iOS app letting users enter a 'virtual room' where notifications and apps are blocked; links to location or time & None\\
Kim et al. 2017 \cite{Kim2017a} & \textit{PomodoLock}, a PC and Android application plus Chrome extension which blocks distracting apps and websites during 25 minute focus sessions & Strength model of self-control \cite{Baumeister2007}\\
Marotta et al. 2017 \cite{Marotta2017} & \textit{Freedom}, a commercial Windows/Mac/Android/iOS app which blocks access to distracting parts of the web or the internet altogether & Rational choice, 'self-commitment' \cite{Bryan2010}\\
Kovacs et al. 2018 \cite{Kovacs2018} & \textit{HabitLab}, a Chrome extension in which the user sets time limit goals for specific sites, then tries a range of interventions to reach the goal & Numerous, including goal setting theory \cite{Locke2002}, operant conditioning \cite{Baron1991}, and self-consistency theory \cite{Sherman1980}\\
Mark et al. 2018 \cite{Mark2018} & \textit{Freedom}, described above & Attentional resources \cite{Wickens1980},  Big 5 \cite{McCrae1999}\\
Okeke et al. 2018 \cite{Okeke2018} & Android app nudging users to close Facebook when a usage limit has been hit, using pulsing vibrations that stop when the user leaves the site & Nudge theory \cite{Sunstein2008}, negative reinforcement \cite{Iwata1987}\\
\bottomrule
\end{tabular}
\end{table*}

\hypertarget{digital-behaviour-change-interventions}{%
\subsubsection{Digital Behaviour Change Interventions}\label{digital-behaviour-change-interventions}}

A large body of HCI work exists on how digital tools can assist
behaviour change in general
\citep{Consolvo2009, Li2010-pers-info, Schueller2013, Kersten-vanDijk2017, Yardley2016, Epstein2015}.
A main focus within such research on `Digital Behaviour Change
Interventions' (DBCIs, \citep{Pinder2018}) is health, for example in relation to how digital
interventions may help users exercise more \citep{Consolvo2009}, quit
smoking \citep{Abroms2011, Heffner2015}, eat more healthily
\citep{Coughlin2015, Okumus2016}, cope with stress
\citep{Konrad2015, Gimpel2015}, or manage chronic conditions \citep{Wang2014}.

Since digital self-control tools can be seen as a subset of DBCIs,
understanding how self-regulation theory has been applied within this
research area is relevant for the present paper. One recent review of 85 DBCI studies \citep{Orji2018}
found the Transtheoretical Model (or Stages of Change) \citep{Prochaska1993} to
be the most commonly referenced (13/85 papers), followed by Goal Setting
Theory \citep{Locke2002} (5/85) and Social Conformity Theory
\citep{Asch1955, Epley1999} (3/85). 60\% (51/85) did not specify any
theoretical basis \citep[cf.][]{Wiafe2012, Schueller2013}, and none specified
dual systems theory \citep[cf.][]{Pinder2018}. The review also found that
among studies which did specify underlying theories, most only
mentioned them without explaining how the theoretical constructs
informed the design and/or evaluation of actual intervention components \citep{Orji2018}.

Another recent comprehensive review \citep{Pinder2018} noted that most
theories applied in DBCI studies assume a rational, deliberative process
as a key determinant of behaviour (e.g.~the Transtheoretical Model
\citep{Prochaska1993} or the Theory of Planned Behaviour \citep{Ajzen1991}).
The authors further argued (after extensive review and discussion) that dual systems theory
could be well placed to guide DBCIs research focusing on long-term
behaviour change through breaking and forming habitual behaviour \citetext{\citealp{Pinder2018, Wood2016, Stawarz2015}; \citealp[cf.][]{Webb2010a}}.

\hypertarget{an-integrative-dual-systems-approach-to-digital-self-control}{%
\section{An Integrative Dual Systems Approach to Digital Self-Control}\label{an-integrative-dual-systems-approach-to-digital-self-control}}

\begin{figure}
\includegraphics[width=1\linewidth,]{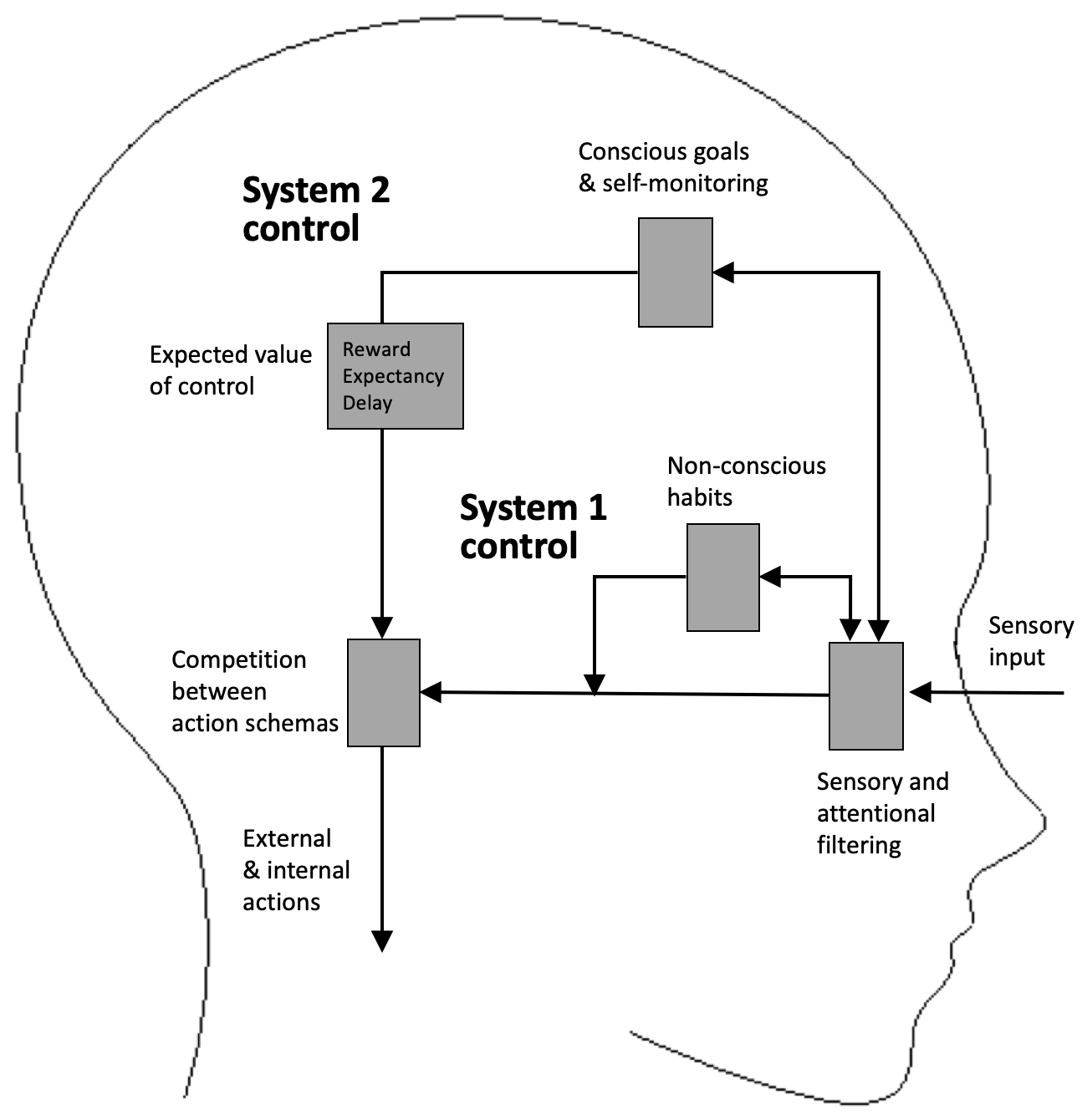}\Description{A flow chart model of self-regulation. System 1 components are i) sensory and attentional filtering and ii) non-conscious habits. System 2 components are i) sensory and attentional filtering, ii) conscious goals and self-monitoring, and iii) expected value of control (with the three subcomponents reward, expectancy, and delay). Both systems then impact on competition between action schemas, the result of which causes behaviour.} \caption{An extended dual systems model of self-regulation, developed from Shea et al. \citep{Shea2014} and Norman \& Shallice \citep{Norman1986}. System 1 control is rapid and non-conscious, whereas System 2 control is slower, conscious, and capacity-limited. The strength of System 2 control is mediated by the expected value of control.}\label{fig:self-regulation-model}
\end{figure}

In the rest of the paper, we draw on dual systems theory and argue how
it can be used to systematise and classify digital self-control tools.
We start by outlining the basics of the underlying psychological
research. In doing so, we extend current applications of dual system
theories in DBCIs \citep{Pinder2018, Cox2016, Pinder2017a, Adams2015} with
the concept of `expected value of control', which the neuroscience
literature considers central in explaining why success at self-control fluctuates over time and with emotional state
\citep{Shenhav2013, Botvinick2015, Inzlicht2014, Ryan2014, Hendershot2011, Tice2001}.
An overview of the resulting model is shown in Figure \ref{fig:self-regulation-model}.

Subsequently, we review current digital self-control tools on the Chrome
Web, Google Play, and Apple App stores, and apply the model to organise
and explain common design features, before pointing out gaps and
opportunities for future work.

\hypertarget{system-1-and-system-2}{%
\subsection{System 1 and System 2}\label{system-1-and-system-2}}

The core of dual systems theories is a major distinction between swift,
parallel and non-conscious `System 1' processes, and slower,
capacity-limited and conscious `System 2' processes
\citep{Kahneman2011, Smith2000, Shiffrin1977, Miller2001, Miller2005, Shea2014, Cooper2014}.\footnote{The related `Nudge theory' \citep{Sunstein2008} draws upon dual systems theories to describe how to exploit System 1 and sometimes System 2 processes to guide people towards a desired action \citep[cf.][]{Adams2015}.}

According to this model, \textbf{System 1} control is driven by environmental
inputs and internal states along with cognitive pathways that map the
current situation to well-learned habits or instinctive responses
\citep{Miller2001}. Behaviour driven by System 1 is often called `automatic',
as it allows tasks to be initiated or performed without conscious
awareness and with little interference with other tasks \citep{Norman1986}.
Instinctive responses like scratching mosquito bites, or frequent
patterns of digital device use like picking up one's smartphone to
check for notifications, can happen automatically via System 1 control
\citep[cf.][]{VanDeursen2015, Oulasvirta2012, Botvinick2015}.

\textbf{System 2} control is driven by goals, intentions, and rules held in
conscious working memory \citep{Miller2005, Baddeley2003}. From these
central representations, signals are sent to cognitive systems that
process sensory input, memory retrieval, emotional processing, and
behavioural output, to guide responses accordingly \citep{Miller2001}. System
2 control is necessary when a goal requires planning or decision-making,
or overcoming of habitual responses or temptations \citep{Norman1986}, for
example if one has a goal of not scratching mosquito bites or not
checking a smartphone notification.

\hypertarget{action-schema-competition}{%
\subsection{Action schema competition}\label{action-schema-competition}}

From a neuroscience perspective, the building blocks of behaviour are
hierarchical \emph{action schemas}, that is, control units for partially
ordered sequences of action that achieve some goal when performed in the
appropriate order \citep{Norman1986, Cooper2000}. Action schemas exist at
varying levels of complexity, from simple single-action motor schemas
for grasping and twisting, to higher level schemas for e.g.~preparing
tea by filling and boiling a kettle \citep{Shallice2011, Botvinick2008}. The
schemas compete for control over behaviour in a `competitive selection'
process in which schemas act like nodes in a network, each with a
continuous activation value \citep{Shallice2011}, and the `winner' is the
node with the strongest activation \citep{Knudsen2007, Pinder2018}. Schema
nodes are activated by a number of sources, including sensory input via
System 1 processes (`bottom-up'), `parental' influence from
super-ordinate schemas in the hierarchy, and top-down influence from
System 2 control \citep{Cooper2000, Shallice2011}.

\hypertarget{self-regulation-and-self-control}{%
\subsection{Self-regulation and self-control}\label{self-regulation-and-self-control}}

Following others, we use \emph{self-regulation} as an umbrella term for
regulatory processes in the service of goal-directed behaviour,
including automatic System 1 habits, and \emph{self-control} more
restrictively for conscious and deliberate System 2 control in
situations where immediate impulses conflict with enduringly valued
goals \citep{Hagger2010, Baumeister2007, Milyavskaya2018, Duckworth2014}.
For example, if a person wishes to be less distracted by her
smartphone in social situations, and through repetition has acquired a
habit of turning the phone face-down to the point that she now does it
without conscious attention, this counts as self-regulation. If in a
given moment she feels an urge to flip it over and check for
notifications, but consciously suppresses this impulse and does not act
on it, this counts as self-control.

Self-regulation and self-control are mediated by feedback functions for
monitoring the state of oneself and the environment, comparing this
state to goals and standards \citep[cf.][]{Shea2014}, and acting to modify the
situation accordingly \citetext{\citealp{Carver1998}; \citealp{Baumeister2000}; \citealp{Inzlicht2014a}; \citealp[cf.~cybernetic models of behaviour control,][]{Carver1981}; \citealp{Powers1973}}.

\hypertarget{attentional-filtering}{%
\subsection{Attentional filtering}\label{attentional-filtering}}

For goals, rules, or intentions to guide System 2 control, they must
first enter working memory \citep{Baddeley2003, Miller2005}. Entry of
information from the external world, internal states, or memory stores
into working memory is itself a competitive process, in which the
signals with the highest activation values are given access by
attentional filters \citep{Constantinidis2007, Miller2001, Knudsen2007}.

Automatic bottom-up filters look out for stimulus properties that are
likely to be important, either through innate sensitivities (e.g.~sudden
or looming noises) or learned associations (e.g.~a smartphone
notification) and boost their signal strength \citep{Knudsen2007}. In this
way, some stimuli may evoke a response strong enough to gain automatic
access to working memory even while we have our minds on other things
\citep{Itti2001, Egeth1997}. For example, clickbait uses headlines and
imagery with properties that makes bottom-up attention filters put its
information on a fast track to conscious working memory or trigger
click-throughs via System 1 control \citep[cf.][]{Blom2015}.

Conscious System 2 control can also direct attention towards particular
internal or external sources of information (e.g.~focusing on a
distorted voice in one's cellphone on a crowded train), which increases
the signal strength of those sources and makes the information they carry
more likely to enter working memory
\citep{Corbetta1991, Muller2005, Knudsen2007, Shu2003}.

\hypertarget{self-control-limitations-and-the-expected-value-of-control}{%
\subsection{Self-control limitations and the Expected Value of Control}\label{self-control-limitations-and-the-expected-value-of-control}}

A central puzzle is why people often fail to act in accordance with
their own valued goals, \emph{even when they are aware of the mismatch}
\citep{Duckworth2016a}. According to current research on cognitive control,
the two key factors to answer this question are (i) limitations on
System 2 control in relation to \emph{capacity}; and (ii) fluctuations due
to \emph{emotional state and fatigue} \citep{Botvinick2015}.

\hypertarget{capacity-limitations}{%
\paragraph{Capacity limitations}\label{capacity-limitations}}
\addcontentsline{toc}{paragraph}{Capacity limitations}

The amount of information that can be held in working memory and guide
System 2 control, is limited (classically `seven, plus or minus two'
chunks of meaningful information \citep{Miller1956, Cowan2010}). Therefore,
self-control can fail if the relevant goals are simply not represented in
working memory at the time of action \citep{Kotabe2015}. This is one
explanation for why people often struggle to manage use of e.g.~Facebook
or email - one opens the application with a particular goal in mind, but
information from the news feed or inbox hijacks attention and crowds out
the initial goal.

\hypertarget{fluctuations-due-to-emotional-state-and-fatigue}{%
\paragraph{Fluctuations due to emotional state and fatigue}\label{fluctuations-due-to-emotional-state-and-fatigue}}
\addcontentsline{toc}{paragraph}{Fluctuations due to emotional state and fatigue}

System 2 control often suffers from fatigue effects if exerted
continuously \citep{Blain2016, Dai2015, Hockey2013, Hagger2010} and also
fluctuates with emotional state \citep{Tice2001}. For example, negative mood
is a strong predictor of relapse of behaviour people attempt to avoid
\citep{Miller1996, Hendershot2011, Tice2001}, and studies of Facebook use
have found that users are worse at regulating the time they spend on the
platform when in a bad mood \citetext{\citealp{Ryan2014}; \citealp[cf.][]{Montag2017}}.

The emerging consensus explanation of these fluctuations is that the
strength of System 2 control is mediated by a cost-benefit analysis of
the outcome that exercising self-control might bring about
\citep{Botvinick2015} i.e., the \textbf{expected value of control} (EVC)
\citep{Shenhav2013}\footnote{The alternative and influential `ego-depletion' explanation
  \citep{Baumeister2007} suggests that System 2 control relies on a limited
  resource that could be depleted, and which has a `refraction period'
  before optimal self-control can again be exercised \citep{Hagger2010}.
  This model has intuitive appeal but has not withstood scrutiny
  \citep{Lurquin2017}, as original studies have failed to replicate
  \citep{Xu2014, Hagger2016}, depletion effects have shown to be
  reversible by increasing the rewards for sustained performance
  \citep{Muraven2003, Hagger2010}, and the purported physical resource
  underlying the effect failed to be discovered
  \citep{Lange2014, Molden2012}.}. The research suggests that EVC is influenced by at
least three major factors:

First, EVC increases the more \textbf{reward} people perceive they could
obtain (or the greater the loss that could be avoided) through
successful self-control
\citep{Botvinick2015, Padmala2010, Padmala2011, Adcock2006}. To
illustrate, consider `phone stack', in which a group dine at a
restaurant and begin by stacking up their phones on the table. The first
person to take out his phone from the stack to check it, has to pay the
entire bill \citep{Ha2012, Tell2013}. This game aids self-control over
device use by introducing a financial (and reputational) cost which
adjusts the overall expected value of control (cf.~also Ko et al.'s tool \emph{NUGU} \citep{Ko2015}).

Second, EVC increases the greater \textbf{expectancy}, or perceived
likelihood, that one will be able to bring a given outcome about through
self-control \citetext{\citealp{Vroom1964}; \citealp{Bernoulli1954}; \citealp{Steel2006}; \citealp[cf. `self-efficacy' in Social Cognitive Theory,][]{Bandura1991}}.
In the phone stack example, people may try harder to suppress an urge to check their phone, the more confidence
they have in their ability to control themselves in the first place.

Third, EVC decreases the longer the \textbf{delay} before the outcome that
self-control might bring about (cf. `future discounting'
\citep{Critchfield2001, Ainslie2001, Evans2016, Ariely2002, McClure2007, ODonoghue2001, Ainslie2010}).
In phone stack, we should expect people to be worse at suppressing an impulse
to check their phone if the rules were changed so that the loser would
pay the bill for a meal in 10 years' time.

\hypertarget{a-practical-example}{%
\subsection{A practical example}\label{a-practical-example}}

As an illustration of the model and the benefits of including EVC,
consider a student who opens his laptop to work on an essay.
However, he instead checks Facebook and spends an inordinate amount of
time scrolling the news feed, experiencing feelings of regret having
done so when he finally returns to the essay. This is not the first time
it happened, even though his reflective goal is always to do solid work
on the essay as the first thing, and to only allow himself to check
Facebook briefly during breaks.

Our model suggests that we think about this situation in terms of the
perceptual cues in the context, automatic System 1 behaviour control,
System 2's consciously held goals and self-monitoring, and System 2's
expected value of control:

If the student normally checks Facebook when opening his laptop, this context may trigger a habitual check-in via System 1 control.
His goal of working first thing might be present in his working memory, but he might fail to override his checking habit due to his expected value of control being low.
This might be because he did not get any \emph{reward} from inhibiting the impulse to check Facebook; because he had little confidence in his own ability to suppress this urge (low \emph{expectancy}); or because the rewards from working on his essay were \emph{delayed} because it was only due in two months.
Alternatively, his goal of working on the essay first thing might not be present in his working memory, in which
case no System 2 control was initiated to override the checking habit.

After having opened Facebook, he might remember that he should be working on the essay, but attention-grabbing content from
the newsfeed enters his capacity-limited working memory and crowd out this goal, leading him to spend more time on Facebook than intended.

\hypertarget{a-review-and-analysis-of-current-digital-self-control-tools}{%
\section{A review and analysis of current digital self-control tools}\label{a-review-and-analysis-of-current-digital-self-control-tools}}

To explore how this model may be useful in mapping digital self-control
interventions, we conducted a systematic review and analysis of apps on
the Google Play and Apple App stores, as well as browser extensions on the
Chrome Web store. We identified apps and browser extensions described as
helping users exercise self-control / avoid distraction / manage
addiction in relation to digital device use, coded their design
features, and mapped them to the components of our dual systems
model\footnote{Data and scripts for reproducing our analyses (as well as this
  paper written in R Markdown \citep[cf.][]{Lyngs2018b}) are available on
  \href{https://osf.io/zyj4h/}{osf.io/zyj4h}.}.

\hypertarget{methods}{%
\subsection{Methods}\label{methods}}

\hypertarget{initial-keyword-search-and-data-clean-up}{%
\subsubsection{Initial Keyword Search and Data Clean Up}\label{initial-keyword-search-and-data-clean-up}}

For the Google Play and Apple App store, we used pre-existing scripts
\citep{Olano2018, Olano2018a} to download search results for the terms
`distraction', `smartphone distraction', `addiction', `smartphone
addiction', `motivation', `smartphone motivation', `self-control' and
`smartphone self-control'. For the Chrome Web store, we developed our
own scraper \citep{Slack2018} and downloaded search results for the same key terms, but
with the prefix `smartphone' changed to `laptop' as well as `internet' (e.g.
`laptop distraction' and `internet distraction'). We separately scraped
apps and extensions on the US and UK stores, between 22nd and 27th August 2018. After excluding duplicate
results returned by multiple search terms and/or by both the US and UK
stores, this resulted in 4890 distinct apps and extensions (1571 from Google Play, 2341 from the App Store, and 978 from the Chrome Web store).

\hypertarget{identifying-potentially-relevant-apps-and-extensions}{%
\subsubsection{Identifying Potentially Relevant Apps and Extensions}\label{identifying-potentially-relevant-apps-and-extensions}}

Following similar reviews \citep{Shen2015, Stawarz2018}, we then manually
screened the titles and short descriptions (if available; otherwise the
first paragraphs of the full description). We included apps and
extensions explicitly designed to help people self-regulate their
digital device use, while excluding tools intended for general
productivity, self-regulation in other domains than digital device use,
or which were not available in English (for detailed exclusion criteria,
see \href{https://osf.io/zyj4h/}{osf.io/zyj4h}).

This resulted in 731 potentially relevant apps and extensions (219 from Google Play, 140 from the App Store, and 372 from the Chrome Web store).

\hypertarget{identifying-apps-and-extensions-to-analyse}{%
\subsubsection{Identifying Apps and Extensions to Analyse}\label{identifying-apps-and-extensions-to-analyse}}

We reviewed the remaining tools in more detail by reading their
full descriptions. If it remained unclear whether an app or extension
should be excluded, we also reviewed its screenshots. If an app existed in both the
Apple App store and the Google Play store, we dropped the
version from the Apple App store.\footnote{Apple's iOS places more restrictions on developer access to
  operating system permissions than does Google's Android, with the
  consequence that the iOS version of a digital self-control app is
  often much more limited than its Android counterpart
  \citep{Mosemghvdlishvili2013}. Because the purpose of our review was
  to investigate which areas of the design space these tools have been
  explored (rather than differences between iOS and Android ecosystems
  per se), we excluded the iOS version when an app was available in
  both stores.}

After this step, we were left with 380 apps and extensions to analyse (96 from Google Play, 60 from the App Store, and 224 from the Chrome Web store).

\begin{figure}
\includegraphics[width=0.98\linewidth,]{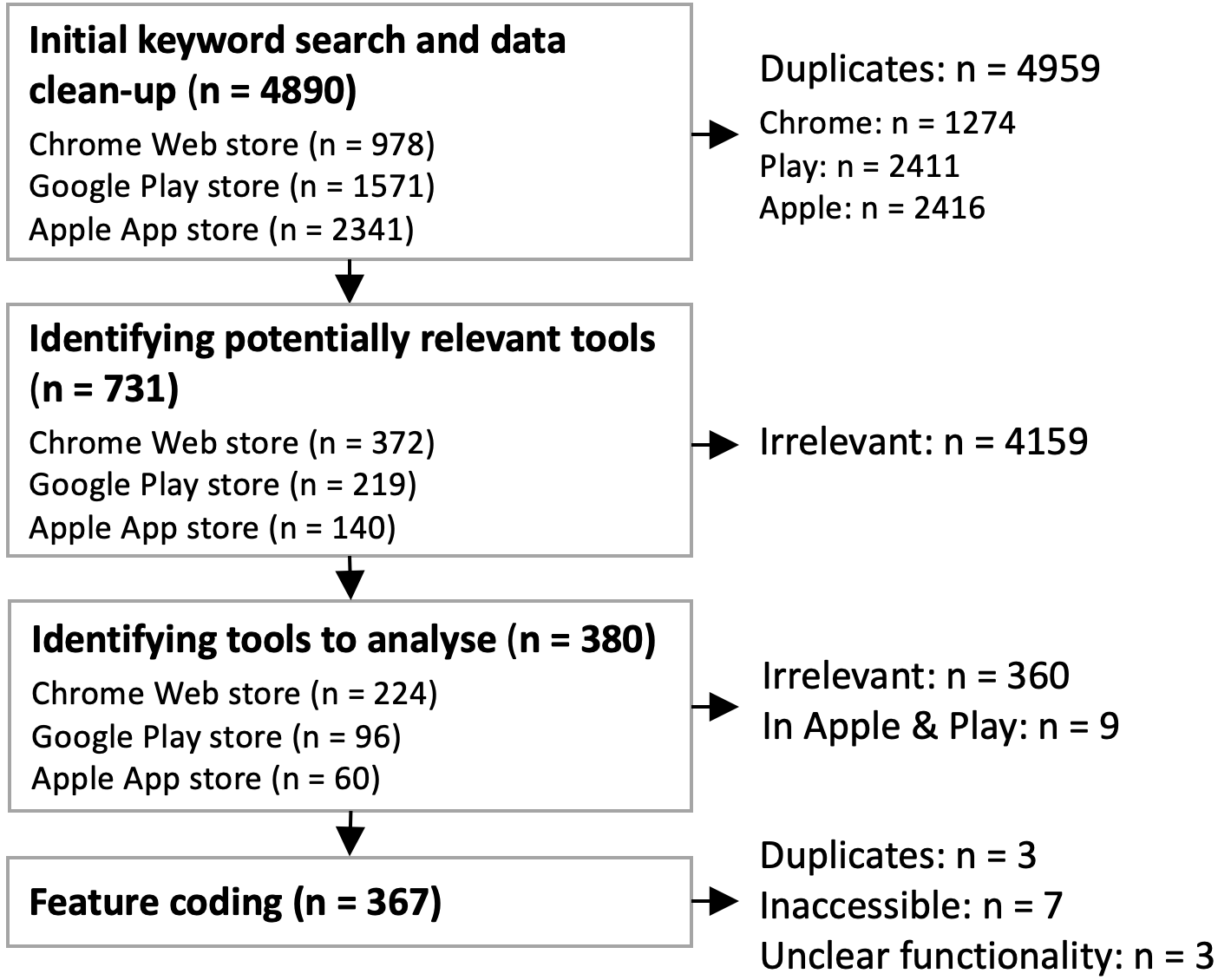}\Description{A flow chart showing the steps in the search and exclusion/inclusion procedure, and how many apps and browser extensions remained after each.} \caption{Flowchart of the search and exclusion/inclusion procedure}\label{fig:inclusion-flowchart}
\end{figure}

\hypertarget{feature-coding}{%
\subsubsection{Feature coding}\label{feature-coding}}

Following similar reviews, we coded functionality based on the
descriptions, screenshots, and videos available on a tool's store page
\citep[cf.][]{Stawarz2014, Stawarz2015, Stawarz2018, Shen2015}. We
iteratively developed a coding sheet of feature categories \citep[cf.][]{Bender2013, Orji2018}, with the prior expectation that the relevant
features would be usefully classified as subcategories of the main
feature clusters `block/removal', `self-tracking', `goal advancement' and
`reward/punishment' (drawing on our previous work in this area \citep{Lyngs2018}).

Initially, three of the authors independently reviewed and classified features in 10 apps and 10 browser extensions (for a total of 30 unique apps and 30 unique browser extensions) before comparing and discussing the feature categories identified and create the first iteration of the coding sheet.
Using this coding sheet, two authors independently reviewed 60 additional apps and browser extensions each and a third author these 120 tools, as well as all remaining.
After comparing and discussing the results, a final codebook was developed, on the basis of which the first author revisited and recoded the features in all tools.
In addition to the granular feature coding, we noted which main feature cluster(s) represented a tool's `core' design, according to the guideline that 25\% or more of the tool's functionality related to that cluster (a single tool could belong to multiple clusters).\footnote{For further detail, see \href{https://osf.io/zyj4h/}{osf.io/zyj4h}.}

During the coding process, we excluded a further 13 tools - 3 duplicates, e.g.~where `pro' and `lite' versions had no difference in described functionality, 7 that had become inaccessible after the initial search, and 3 that lacked sufficiently well-described functionality to be coded. This left 367 tools in the final dataset.

\hypertarget{results}{%
\subsection{Results}\label{results}}

\hypertarget{feature-prevalence}{%
\subsubsection{Feature prevalence}\label{feature-prevalence}}

\begin{figure}
\includegraphics[width=0.98\linewidth,]{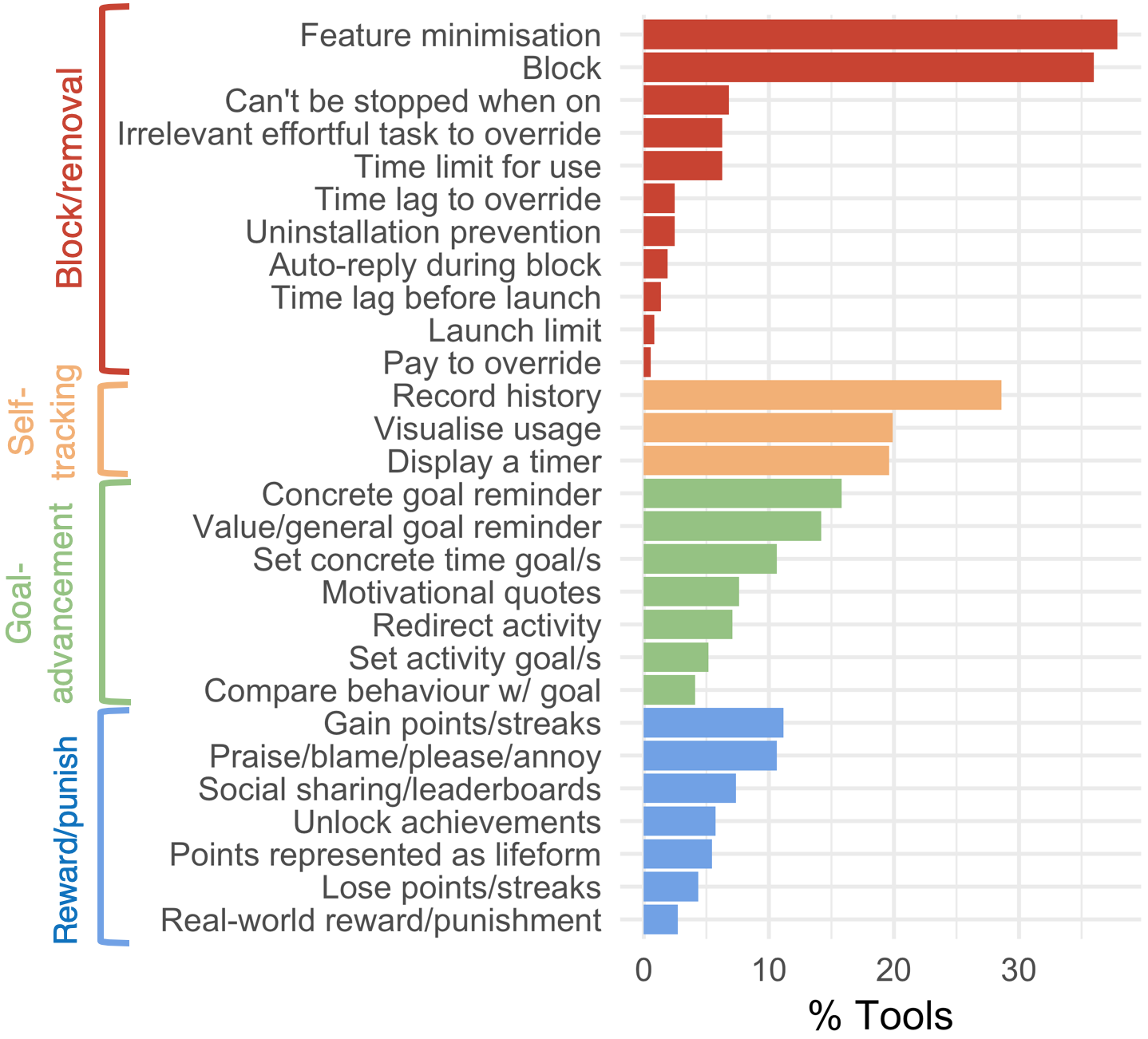}\Description{A grouped bar chart which shows the percentage of tools that incorporated specific elements relating to the feature clusters block/removal (11 features shown), self-tracking (3 features shown), goal-advancement (7 features shown), and reward/punishment (7 features shown).} \caption{Functionality of digital self-control tools (N = 367)}\label{fig:feature-frequency}
\end{figure}

A summary of the prevalence of features is shown in Figure \ref{fig:feature-frequency}.
The most frequent feature cluster related to \textbf{blocking or removing} distractions, some variation of which was present in 74\% of tools.
44\% (163) enabled the user to put obstacles in the way of distracting functionality by either blocking access entirely (132 tools), or by setting limits on how much time could be spent (23 tools) or how many times distracting functionality could be launched (3 tools) before being blocked, or by adding a time lag before distracting functionality would load (5 tools).
14\% of tools (50) also added friction if the user attempted to remove the blocking, including disallowing a blocking session from being stopped (25 tools), requiring the user to first complete an irrelevant effortful task or type in a password (23 tools), tinkering with administrator permissions to prevent the tool from being uninstalled (9 tools), or adding a time lag before the user could override blocking or change settings (9 tools).
For example, the \emph{Focusly} Chrome extension \citep{Trevorscandalios2018} blocks sites on a blacklist; if the
user wishes to override the blocking, she must type in correctly a series of 46 arrow keys (e.g. \(\rightarrow\uparrow\downarrow\rightarrow\leftarrow\rightarrow\)...) correctly to enter the blocked site.

Rather than blocking content per se, an alternative approach, taken by 38\% of tools (139), was to reduce the user's exposure to distracting options in the first place.
This approach was dominated by browser extensions (121 of these tools were from the Chrome Web store) typically in the form of removing elements from specific sites (67 tools;
e.g.~removing newsfeeds from social media sites or hiding an email inbox).
The sites most frequently targeted were Facebook (26 tools),
YouTube (17), Twitter (11) and Gmail (7).
Also popular were general `reader' extensions for removing distracting content when browsing the
web (27 tools) or when opening new tabs (24).
Other notable examples were `minimal-writing' tools (22 tools) which remove
functionality irrelevant to, or distracting from, the task of writing.
Finally, a few Android apps (4 tools) limited the amount of functionality available on devices' home screen.

The second most prevalent feature cluster related to \textbf{self-tracking}, some variation of which was present in 38\% of tools (139).
Out of these, 105 tools recorded the user's history, 73 provided visualisations of the
captured data, and 72 displayed a timer or countdown.
In 42 tools, the self-tracking features included focused on the time during which the user managed to \emph{not} use their digital devices, such as the iOS app \emph{Checkout of your phone} \citep{Schungel2018}.

The third most prevalent feature cluster related to \textbf{goal advancement}, some variation of which was present in 35\% of tools (130).
58 tools implemented reminders of a concrete time goal or task the user tried to complete (e.g.~displaying pop-ups when a set amount of time has been spent on a distracting site or by replacing the content of newsfeeds or new tabs with todo-lists) and 52 tools provided reminders of more general goals or personal values (e.g.~in the form of motivational quotes).
58 tools asked the user to set explicit goals, either for how much time they wanted to spend using their devices in total or in specific apps or websites (39 tools), or
for the tasks they wanted to focus on during use (19 tools). 15 tools allowed the user to compare their actual behaviour against the goals they set.

The fourth most prevalent feature cluster, present in 22\% of tools (80), related to \textbf{reward/punishment}, i.e.~providing some rewards or punishments for the way in which a device is used.
Some of these features were gamification interventions such as collecting points/streaks (41 tools), leaderboards or social sharing (27), or
unlocking of achievements (21).
In 20 tools, points were represented as some lifeform (e.g.~an animated goat or a growing tree) which might be harmed if the user spent too much time on certain websites or used their phone during specific times.
10 tools added real-world rewards or punishments, e.g.~making the user lose money if they spend more than 1 hour on Facebook in a day (\emph{Timewaste Timer} \citep{Prettymind.co2018}), allowing virtual points to be exchanged to free coffee
or shopping discounts (\emph{MILK} \citep{MilkTheMomentInc.2018}) or even let the user administer herself electrical shocks when accessing blacklisted websites (!) (\emph{PAVLOK} \citep{Pavlok2018}).

Finally, 35\% of tools (129) gave the user control over what counted as `distraction', e.g.~by letting the user customise which apps or which websites to restrict access to. Among tools implementing blocking functionality, 101 tools implemented blacklists (i.e.~blocking specific apps or sites, allowing everything else), while 22 tools implemented whitelists (i.e.~allowing only specified apps or sites while blocking everything else).

\hypertarget{feature-combinations}{%
\subsubsection{Feature combinations}\label{feature-combinations}}

65\% of tools had only one feature cluster at the core of their design, the most frequent of which was blocking/removing distractions (53\%).
32\% (117 tools) combined two main feature clusters, most frequently
block/removal in combination with goal-advancement (40 tools; e.g.~replacing the Facebook newsfeed with a todo list, or replacing distracting websites with a reminder of the task to be achieved) or self-tracking in combination with reward/punishment features (30 tools; e.g.~a gamified pomodoro timer in which an animated creature dies if the user leaves the app before the timer runs out).
Block/removal core designs were also commonly combined with self-tracking (24 tools; e.g.~blocking distracting websites while a timer counts down, or recording and displaying how many times during a block session the user tried to access blacklisted apps).
Only two tools (\emph{Flipd} \citep{FlipdInc2018} and \emph{HabitLab} \citep{habitlab-extension}) combined all four feature clusters in their core design, with the Chrome extension HabitLab (developed by the Stanford HCI Group) cycling through different types of interventions to learn which best help the user align internet use with their stated goals \citep[cf.][]{Kovacs2018}.

\hypertarget{store-comparison}{%
\subsubsection{Store comparison}\label{store-comparison}}

\begin{figure}
\includegraphics[width=0.98\linewidth,]{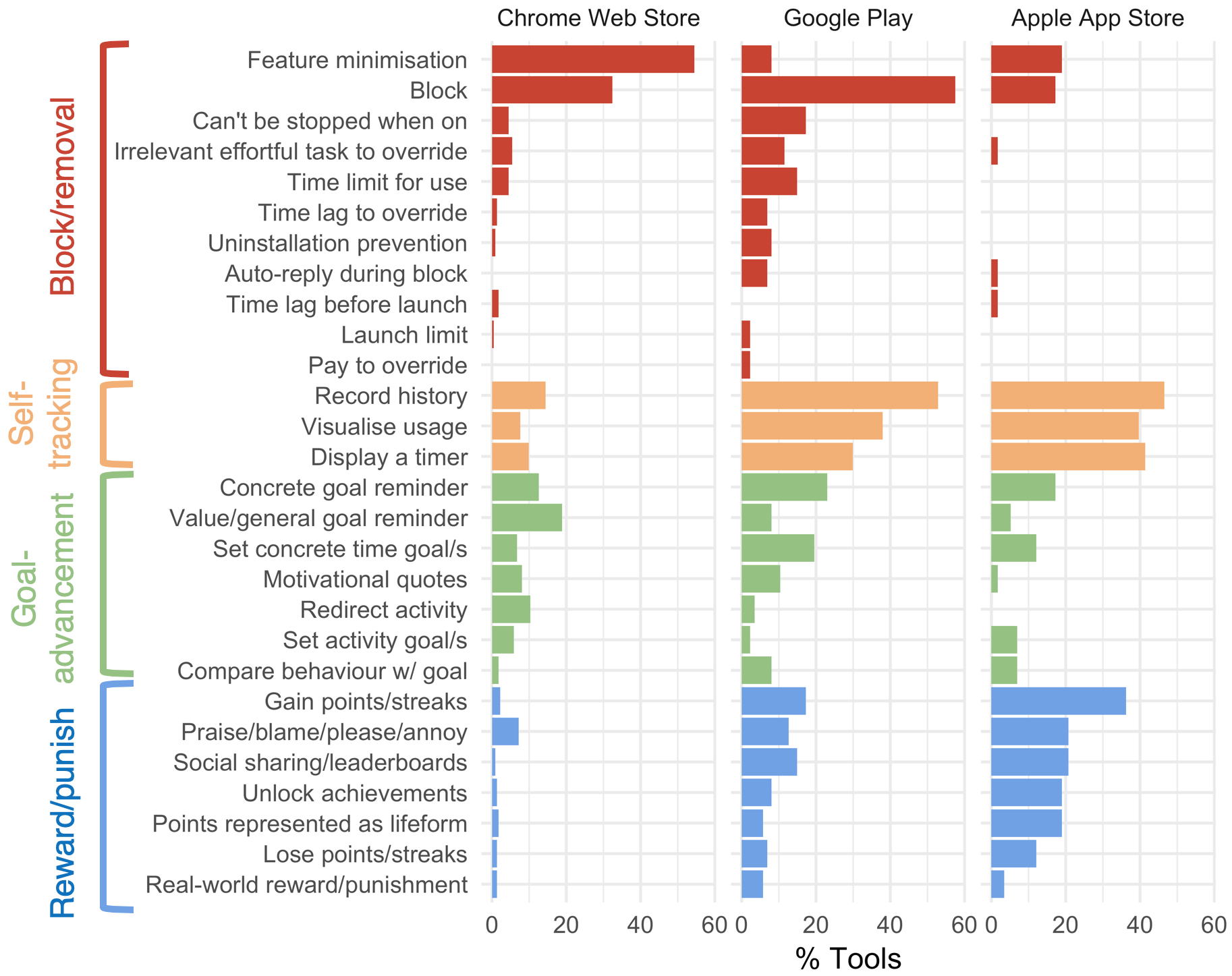}\Description{A grouped bar chart of feature frequency, broken down by store. That is, it shows separately for the Chrome Web, Google Play, and Apple App store the percentage of tools that incorporated specific elements relating to the feature clusters block/removal (11 features shown), self-tracking (3 features shown), goal-advancement (7 features shown), and reward/punishment (7 features shown).} \caption{Functionality of digital self-control tools on Chrome Web (n = 223), Google Play (n = 86) and Apple App Store (n = 58)}\label{fig:compare-features-in-stores}
\end{figure}

Figure \ref{fig:compare-features-in-stores} summarises the prevalence of features, comparing the three stores.
The differences between the stores appear to mirror the granularity of system control available to developers: Feature minimisation, in the form of removing particular aspects of the user interface, is common in browser extensions, presumably because developers here can wield precise control over the elements displayed on HTML pages by injecting client-side CSS and JavaScript.
On mobile devices, however, developers have little control over how another app is displayed, leaving blocking or restricting access as the only viable strategies.
The differences between the Android and iOS ecosystems are apparent, as the permissions necessary to implement e.g.~scheduled blocking of apps are not available to iOS developers.
These differences across stores suggest that if mobile operating systems granted more permissions (as some developers of popular anti-distraction tools have petitioned Apple to do \citep{Warriors2018}), developers would respond by creating tools that offer more granular control of the mobile user interface, similar to those that already exist for the Chrome web browser.

\hypertarget{mapping-identified-tool-features-to-theory}{%
\subsubsection{Mapping identified tool features to theory}\label{mapping-identified-tool-features-to-theory}}

\begin{figure}
\includegraphics[width=0.98\linewidth,]{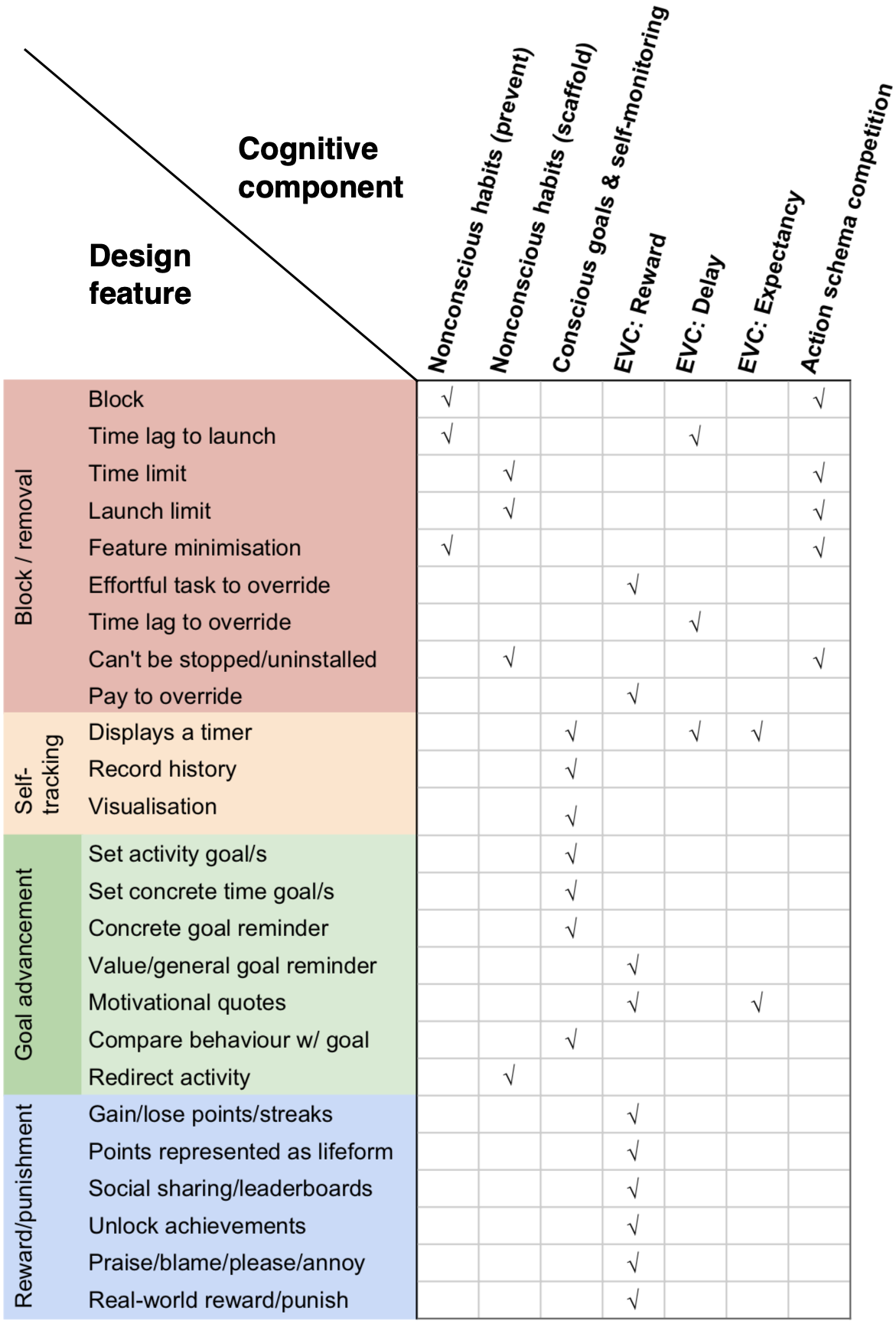}\Description{Matrix which for each of the 25 design features displays which components of the dual systems model they have the most direct potential to influence} \caption{Mapping of design features to an integrative dual systems model of self-regulation}\label{fig:feature-to-theory-mapping}
\end{figure}

Figure \ref{fig:feature-to-theory-mapping} shows a matrix of how the
design features corresponds to the main components of the integrative
dual systems model, in terms of the cognitive components they have the
most immediate potential to influence:
\emph{Non-conscious habits} are influenced by features that block the targets of habitual action
or the user interface elements that trigger them, thereby preventing unwanted habits from being activated.
\emph{Non-conscious habits} are also influenced by features which enforce limits on daily use, or redirect user activity, thereby scaffolding formation of new habits.
\emph{Conscious goals \& self-monitoring} is influenced by explicit goal setting and reminders,
as well as by timing, recording, and visualising usage and comparing it
with one's goals.
The \emph{reward} component of the expected value of control is influenced by reward/punishment features that add incentives
for exercising self-control, as well as by value/general goal reminders and motivational
quotes which encourage the user to reappraise the value of
immediate device use in light of what matters in their life; the
\emph{delay} component is influenced by time lags or timers; and
\emph{expectancy} is similarly influenced by timers (`I should be able to manage
to control myself for just 20 minutes!') as well as motivational quotes.
Finally, the \emph{action schema competition}, which ultimately controls behaviour, is most directly
affected by blocking/removal functionality that hinders unwanted responses
from being expressed by simply making them unavailable.

\begin{figure}
\includegraphics[width=0.98\linewidth,]{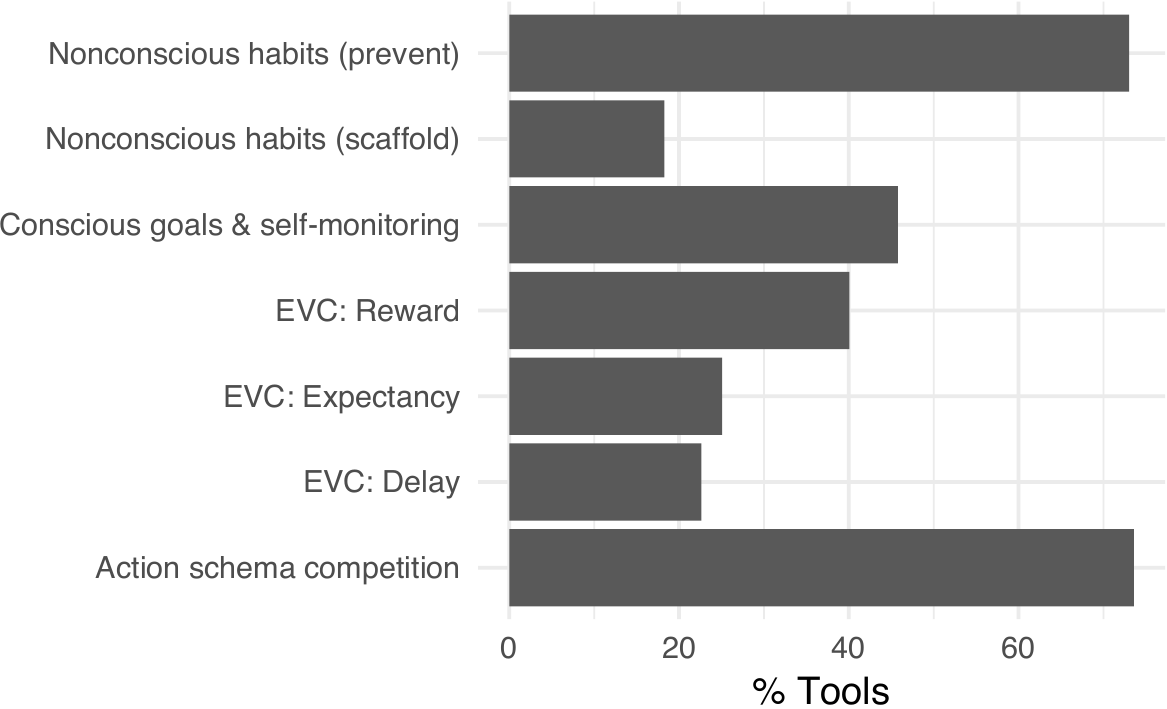}\Description{A bar chart which for each cognitive component of the dual systems self-regulation model shows the percentage of tools that include at least one design feature targeting it.} \caption{Percentage of tools which include at least one design feature targeting a given cognitive component of the dual systems model of self-regulation.}\label{fig:cognitive-target-frequency}
\end{figure}

Given this mapping, the percentages of tools in which at least one design feature maps to a given cognitive component is shown in Figure \ref{fig:cognitive-target-frequency}.
Similarly to DBCI reviews \citep{Pinder2018, Stawarz2015}, we find the lowest prevalence of features that scaffold formation of non-conscious habits (18\%), followed by
features that influence the delay component of the expected value of control (23\%).
The current landscape of digital self-control tools is dominated by features which prevent activation of unwanted non-conscious
habits (73\%) and thereby stop undesirable responses from winning out in action schema competition by making them unavailable.

\hypertarget{discussion}{%
\section{Discussion}\label{discussion}}

This study set out to map the landspace of current digital self-control
tools and relate them to an integrative dual systems model of
self-regulation. Our review of 367 apps and browser extensions found
that blocking distractions or removing user interface features were the most common approaches to digital self-control.
Grouping design features into clusters, the prevalence ranking was block/removal \textgreater{} self-tracking \textgreater{} goal advancement \textgreater{} reward/punishment.
Out of these, 65\% of tools focused on only one
cluster in their core design; and most others (32\%) on two.
The frequencies of design features differed between the Chrome Web Store, Play Store, and Apple App store, which likely reflects differences in developer permissions.
When mapping design features to our dual systems model, the least commonly targeted cognitive component was unconscious habit scaffolding, followed by the delay and expectancy elements of the expected value of control.

We now turn to discuss how these empirical observations can inform future research by pointing to: i) widely used and/or theoretically interesting design features in current digital self-control tools that are underexplored in HCI research; ii) feature gaps identified by our application of the dual systems model, showing neglected areas that could be relevant to researchers and designers, and iii) how the model may be used directly to guide research and intervention design. We then outline limitations and future work.

\hypertarget{research-opportunities-prompted-by-widely-used-or-theoretically-interesting-design-features}{%
\subsection{Research opportunities prompted by widely used or theoretically interesting design features}\label{research-opportunities-prompted-by-widely-used-or-theoretically-interesting-design-features}}

The market for digital self-control effectively amounts to hundreds of natural experiments in supporting self-control, meaning that successful tools may reveal design approaches with wider applicability.
These approaches present low-hanging fruit for research studies, especially as many are so far lacking evaluation in terms of their efficacy and the transferability of their underlying design mechanisms.
As an example, we highlight three such instances:

\emph{Responsibility for a virtual creature}: \emph{Forest} \citep{Forestapp.cc2018} ties device use to the well-being of a virtual tree.
Numerous variations and clones of this approach exist among the
tools reviewed, but Forest is the most popular with over 5 million users
on Android alone. It presents a novel use of `virtual pets' that
requires the user to abstain from action (resist using their phone)
rather than take action to `feed' the pet, and is a seemingly successful example of
influencing the reward component of expected value of control.

\emph{Redirection of activity}: \emph{Timewarp} \citep{Stringinternational.com2018}
reroutes the user to a website aligned with their productivity goals
when navigating to a distracting site (e.g.~from Reddit to Trello), and
numerous tools implement similar functionality. Such apps seem to
be automating `implementation intentions' (\emph{if-when} rules for
linking a context to a desired response \citep{Gollwitzer2006}), an intervention which
digital behaviour change researchers have highlighted as a promising way to
scaffold transfer of conscious System 2 goals to automatic System 1
habits \citep{Pinder2018, Stawarz2015}.

\emph{Friction to override past preference}: A significant number of tools
not only allow the user to restrict access to digital distractions,
but also add a second layer of commitment, e.g.~by making blocking
difficult to override, as in the browser extension \emph{Focusly} \citep{Trevorscandalios2018}, which requires a laborious combination of keystrokes
to be turned off. This raises important design and ethical
questions about how far a digital tool should go to hold users
accountable for their past preferences \citep[cf.][]{Bryan2010, Lyngs2018a}.

\hypertarget{gaps-identified-by-the-dual-systems-model}{%
\subsection{Gaps identified by the dual systems model}\label{gaps-identified-by-the-dual-systems-model}}

By applying our model, we also identified three cognitive mechanisms that appear underexplored by current digital self-control tools.
We argue that focusing on these mechanisms could lead to new powerful models for digital self-control:

\hypertarget{scaffolding-habits}{%
\paragraph{Scaffolding habits}\label{scaffolding-habits}}
\addcontentsline{toc}{paragraph}{Scaffolding habits}

Similar to the situation in general DBCIs \citep[cf.][]{Pinder2018, Stawarz2015}, the least frequently targeted cognitive component relates to \emph{scaffolding of new, desirable unconscious habits} (as opposed to preventing undesired ones from being triggered via blocking or feature removal).
Habit formation is crucial for long-time behaviour change, and in the context of DBCIs, Pinder et al. \cite{Pinder2018} suggested implementation intentions and automation of self-control as good candidate strategies for habit targeting. We note that some such design interventions are already being explored amongst current digital self-control tools: Apart from the tools mentioned above that redirect activity, we highlight that four tools allow blocking functionality to be linked to the user's location (e.g. \textit{AppBlock} \cite{MobileSoft2019} and \textit{Bashful} \cite{Runnably2019}). We expect this to be a powerful way of automatically triggering a target behaviour in a desired context.

\hypertarget{delay}{%
\paragraph{Delay}\label{delay}}
\addcontentsline{toc}{paragraph}{Delay}

The \emph{delay} component of expected value of control is also less commonly targeted:
the number of tools including functionality targeting delay drops to 4\% if we exclude the display of a timer (which raises time awareness rather than affecting actual delays).
This is surprising from a theoretical perspective, because the effects on behaviour of sensitivity to delay are strong, reliable, and---at least to behavioural economists---at the core of self-control difficulties \citep{Ariely2002, Dolan2012}.
Even if rewards introduced by gamification features may have the side effect of reducing delay before self-control is rewarded, it remains surprising that only two of 367 reviewed tools directly focused on using delays to scaffold successful
self-control (\emph{Space} \citep{BoundlessMindInc2018} increases launch times for distracting apps on iOS; \emph{Pipe Clogger} \citep{Croshan2018} does the same for websites). As previous research has found people to be especially sensitive to delays in online contexts \citep{Krishnan2013}, we expect interventions
that leverage delays to scaffold self-control in digital environments to be highly effective.

\hypertarget{expectancy}{%
\paragraph{Expectancy}\label{expectancy}}
\addcontentsline{toc}{paragraph}{Expectancy}

The \emph{expectancy} component (i.e.~how likely a user think it is
that she will be able to reach her goal through self-control exertion)
was also less frequently targeted, and mainly through timers limiting the
duration where the user tried to exert self-control. Given the crucial
role of self-efficacy in Bandura's influential work on self-regulation
\citep{Bandura1991}, this may also represent an important underexplored
area. One interesting approach to explore is found in \emph{Wormhole Escaper}
\citep{Bennett2018} which lets the user administer words of encouragement to
themselves when they manage to suppress an urge to visit a distracting
website. In so far as this is effective, it may be by
boosting the user's confidence in their ability to exert self-control.

\hypertarget{using-the-model-directly-to-guide-intervention-research-and-design}{%
\subsection{Using the model directly to guide intervention research and design}\label{using-the-model-directly-to-guide-intervention-research-and-design}}

The abstracted nature of the model enables it to be utilised on different levels of analysis to inspire new avenues for research as well as drive specific design:

For researchers, the model may be used to organise existing work on design interventions by the cognitive components targeted, as well as a roadmap for future studies that focus on different components of the self-regulatory system.
Whereas many other theories and frameworks are on offer for this purpose, one advantage of the dual systems model is that it provides HCI researchers with clear connections to wider psychological research on basic mechanisms of self-regulation, which can be utilised in design.

As such, the model may be used as a starting point for design consideration that is aligned with the cognitive mechanisms involved in self-regulation; its components can be readily expanded if inspiration from more theoretical details and predictions is required. For example, the `reward' component readily expands into more specific models explaining the types of stimuli that may be processed as rewards; how timing of rewards impact their influence; how the impact of gains differ from losses; and so on \citep{Berridge2015, Caraco1980, Schull2012, Kahneman1979}.

Two recent examples in HCI research illustrate such possible use of psychological theory in the design process:
In the design and development of \emph{TimeAware}, Kim et al.~were guided by work on differential sensitivity to gains vs losses, and found that their visualisation tool more effectively supported productivity when displaying time spent engaging with distracting rather than productive activities \citep{Kim2016}. Similarly, based on dual systems theory, Adams et al. \citep{Adams2015} trialed ways of applying visual and auditory perception biases in design interventions to influence food choice and voice pitch.
We hope our model may inspire designs that are similarly informed by psychological theory.

\hypertarget{limitations-and-future-work}{%
\subsection{Limitations and future work}\label{limitations-and-future-work}}

Our review has some limitations. First, due to space restrictions, and
because information about numbers of users are not available on the
Apple App Store, our tool analysis has focused on functionality
analysis, while leaving consideration of numbers of installs or content
of user reviews to future work. We note that this is similar to the
approach taken by other reviews in related areas
\citep{Stawarz2015, Stawarz2014}.

Second, the integrative dual systems model we have applied points to
directions of future research, but its high-level formulation
leaves its cognitive design space under-specified. How precisely one
should be able to anchor details of specific interventions directly in
causal theories is a point of longstanding debate \citep{Michie2008, Hardeman2005, Ajzen1991}.
A main benefit of dual systems theory, however, is that while concise, it remains directly grounded in
well-established basic research on self-regulation.
As mentioned above, this means that each component of the model has substantial literature behind it, so that more detailed specifications and predictions can be found in lower-level theories on demand.

Turning to the future, self-control in relation to digital device use
involves unique challenges and opportunities compared to general
behaviour change research. On the one hand, portable, powerful,
internet-connected devices present an unprecedented self-regulation challenge:
Never before have so many behavioural options,
information about nearly everything, engaging games, and communication
with friends, family, and strangers, been instantly available.
On the other hand, this very challenge presents a unique research opportunity.
Precisely \emph{because} digital devices afford so much functionality, they allow us to
test design interventions with greater precision,
flexibility, and dramatically lower cost than changing the physical
environment. Moreover, context detection, a constant challenge in DBCI
research for administering meaningful and well-timed interventions
\citep{Pinder2018}, is more manageable in relation to device use, because a large
amount of relevant activity can be easily measured.

The research on digital self-control tools should therefore be of wide interest as a test bed for interventions that optimise self-control in an environment where most factors can be changed at minimal cost.

\hypertarget{conclusion}{%
\section{Conclusion}\label{conclusion}}

The challenge of designing powerful, always-connected digital devices that support self-control over their use, is important to address.
This paper contributes to such efforts on two levels: (i) by providing the first
comprehensive functionality analysis of current apps and browser
extensions for digital self-control on the Google Play, Chrome Web, and
Apple App stores, and (ii) by applying a well-established model of
self-regulation to evaluate their design features and provide a
mechanistic understanding of the problem they address.

The future to hope for is one in which users develop beneficial habits of
technology use and are resilient against predatorial nudging by
clickbait advertisers and data harvesters. We hope our review of 367 apps and browser extensions representing natural experiments in designing for digital self-control, and our formulation of a dual
systems model to understand them, will help us realise this future.

\bibliographystyle{ACM-Reference-Format}
\balance
\bibliography{references.bib}

\end{document}